\documentstyle [fleqn,12pt]{article}
\def\x{\phantom{$-$}}

\def\thalf{{\textstyle{\frac{1}{2}}}}
\def\pj{\hspace{-0.27cm}}
\def\qh{\hat{Q}}
\def\pal{P_{\alpha}}
\def\pab{P_{\beta}}
\def\epa{\epsilon_{\alpha}}
\def\epb{\epsilon_{\beta}}
\def\epc{\epsilon_{\gamma}}
\def\dw{\delta\omega}
\begin{document}
\begin{titlepage}
\pagestyle{empty}
\begin{center}
\begin{large}
{{\bf Iterative Solution for Effective Interactions in a System with
Non-degenerate Unperturbed Energies}}
\end{large}
\vskip 1cm
K. Suzuki$^{a}$, R. Okamoto$^{a}$, P.J. Ellis$^{b}$,
and T.T.S. Kuo$^{c}$\\
\vskip 1cm
\begin{em}{\small
$^a$Department of Physics, Kyushu Institute of Technology, Kitakyushu 804,
Japan}\\
\end{em}
\begin{em}{\small
$^b$School of Physics and Astronomy, University of Minnesota, Minneapolis,
MN 55455}\\
\end{em}
\begin{em}{\small
$^c$Department of Physics, State University of New York at Stony Brook,
Stony Brook, NY 11794}\\
\end{em}
\vskip 0.7cm
{\bf Abstract}
\end{center}
We generalize the Lee-Suzuki iteration method for summing the folded
diagram series to the case where the unperturbed model-space energies are
non-degenerate. A condition is derived for the convergence of the iteration
scheme and this depends on the choice of the model space projection operators.
Two choices are examined, in the first the projection operators are defined in
terms of the unperturbed states and in the second they are defined in terms of
the eigenfunctions obtained at each stage of the iteration. As is illustrated
by calculations with a simple model, the second procedure gives the better
convergence and, by suitable choice of the starting energies, allows the
reproduction of any subset of the exact eigenvalues.
\vskip0.7cm
\noindent PACS number: 21.60.Cs
\vskip-22.5cm
\phantom{1234}\hfill NUC-MINN-93/11-T
\end{titlepage}
\section{Introduction}

In nuclear, atomic and chemical physics it is usually necessary
to recast the full many-body problem in the form of an effective interaction
acting  within a chosen model space for which the eigenvalues can be obtained
exactly. Much work has been carried out on this topic, both as regards formal
questions and actual calculations, see refs. [1--5] for the nuclear case.
The formalism consists of a completely linked perturbation series which
contains both non-folded and folded diagrams. For a given set of non-folded
diagrams, the folded diagram series can be summed by using either the
Krenciglowa-Kuo (KK) technique \cite{emt} or the Lee-Suzuki (LS) method
\cite{lees}. Both these methods employ  completely degenerate unperturbed
energies for the model space, however their convergence properties
are different.
The KK approach, when convergent, yields the eigenvalues for those states
which have the largest overlap with the chosen model space. On the other hand
the LS method reproduces those eigenvalues which lie closest to the chosen
unperturbed energy.

Since actual single particle energies in, for example the $(sd)$ or the $(pf)$
shells, are far from degenerate, it is clearly desirable to use a formalism
which is not restricted to exact degeneracy. This would allow one to treat the
one-body terms as unperturbed energies rather than introducing artificial
energy
shifts so as to rewrite the problem in degenerate form. Further the KK and
LS methods yield only certain of the exact eigenvalues which is not, in
general, desirable. This restriction is not present in the exact
representation of the complete many-body problem as a
series of non-folded and folded diagrams since this contains information
regarding all of the true eigenvalues. Given a model space of dimension $d$ it
should be possible to obtain any selection of $d$ eigenvalues from the
complete set of true eigenvalues. We shall show that this is indeed possible
using the formalism developed in this paper which is expressly designed for
non-degenerate unperturbed energies.

The organization of this paper is as follows. In sec. 2 we briefly outline the
standard LS formalism and establish notation. The generalization of the LS
approach to the non-degenerate case is discussed in sec. 3. Since the solution
of the equations requires iteration, we give in sec. 4 a criterion for the
convergence of the iteration. In order to assess the present approach and to
compare with the KK and LS methods, we need to study a case where the exact
results are known, i.e. a model. This is the subject of sec. 5. Our
concluding remarks are given in sec. 6.

\section{Outline of the Lee-Suzuki Method}

In the usual way we write the full Hamiltonian, $H=H_0+V$, where $H_0$ is the
unperturbed Hamiltonian and $V$ is the perturbation. We use a basis in which
$H_0$ is diagonal and define an
operator $P$ which projects the Hilbert space onto the chosen model space of
dimension $d$.
The complementary operator $Q$ projects onto the remainder of the Hilbert
space, thus $P+Q=1,\ PQ=0$. Then it is straightforward to obtain formal
expressions for the standard non-Hermitian effective interaction, which we
denote here by $R$. It can be derived by a number of different methods
\cite{bhb,tto,suz} and can be written in various forms. Here it is
convenient to use
\begin{equation}
R=PVP+PVQ\omega\;,
\end{equation}
where the operator $\omega$ obeys the equation
\begin{equation}
QVP+QHQ\omega-\omega PHP-\omega PVQ\omega=0\;.
\end{equation}
Then the model-space eigenvalue equation, yielding $d$ of the true
eigenvalues labelled $E_p$, can be written
\begin{equation}
(PH_0P+R)|\phi_p\rangle=E_p|\phi_p\rangle\;,
\end{equation}
where the model space wave function is the projection of the true wave function
on the model space, {\it i.e.}, $\phi_p=P\Psi_p$.

In this section we consider a system with degenerate unperturbed energies, thus
\begin{equation}
PH_0P=\epsilon_0P\;.
\end{equation}
In this case the equation for $\omega$ becomes
\begin{equation}
(\epsilon_0-QHQ)\omega+\omega(PVP+PVQ\omega)=QVP\;,
\end{equation}
which can be rewritten in the form
\begin{equation}
\omega=\frac{1}{\epsilon_0-QHQ}QVP-\frac{1}{\epsilon_0-QHQ}\omega R\;.
\end{equation}
Substituting eq. (6) into eq. (1), we have
\begin{equation}
R=\left[1+PVQ\frac{1}{\epsilon_0-QHQ}\,\omega\right]^{-1}\hat{Q}(\epsilon_0)\;,
\end{equation}
where we have defined the $\hat{Q}$-box according to
\begin{equation}
\hat{Q}(\epsilon_0)=PVP+PVQ\frac{1}{\epsilon_0-QHQ}QVP\;.
\end{equation}

The LS method \cite{lees} generates the solution of eqs. (6) and (7)
by iteration. The $n^{\rm th}$ iteration is obtained from the $(n-1)^{\rm st}$
iteration by the following equations
\begin{eqnarray}
R_n&\pj=&\pj\left[P+PVQ\frac{1}{\epsilon_0-QHQ}\,\omega_{n-1}\right]^{-1}\hat{Q}
(\epsilon_0)\;,\\
{\rm and}&&\nonumber\\
\omega_n&\pj=&\pj\frac{1}{\epsilon_0-QHQ}QVP-\frac{1}{\epsilon_0-QHQ}
\,\omega_{n-1}R_n\;.
\end{eqnarray}
Defining $\omega_0=0$, the $n^{\rm th}$ iterative solution, ($n>2$), is
given by
\begin{eqnarray}
R_n&\pj=&\pj\left[P-\qh_1-\qh_2R_{n-1}-\cdots-\qh_{n-1}R_2R_3\cdots R_{n-1}
\right]^{-1}\qh\nonumber\\
&\pj=&\pj\left[P-\qh_1-\sum\limits_{m=2}^{n-1}\qh_m
\prod\limits_{k=n-m+1}^{n-1}R_k\right]^{-1}\qh\;,
\end{eqnarray}
where $\qh_m$ for $m=1,2,\ldots$ is given by the $m^{\rm th}$ derivative of
the $\qh$-box, namely
\begin{equation}
\qh_m=(-1)^mPVQ\left(\frac{1}{\epsilon_0-QHQ}\right)^{\!\!m+1}QVP
=\frac{1}{m!}\frac{d^m\qh(\epsilon_0)}{d\epsilon_0^m}\;.
\end{equation}
The sequence $\{R_1,R_2,\ldots\}$ is generally convergent (see sec. 4) so the
solution for the effective interaction corresponds to $R=R_{\infty}$. The
rate of convergence and, indeed, which of the exact eigenvalues are obtained
from $R_{\infty}$ will depend on the unperturbed energy $\epsilon_0$. It
is important to realize that this is a parameter at our disposal. Thus the
Hamiltonian and eq. (5) are unchanged if we shift $\epsilon_0$ to
$\epsilon_0+\epsilon'$ and compensate for this with
a corresponding shift of the perturbation $V$ to $V-\epsilon'P$; the quantity
$\epsilon'$ is clearly arbitrary and can be used to optimize the convergence
of the iteration procedure.

\section{The Non-degenerate Case}
\subsection{General Iterative Solution for Effective \protect\\Interactions}
Let us define projection operators, $\pal$, which act in the model space
and are such that
\begin{equation}
P=\sum\limits_{\alpha}\pal\qquad{\rm and}\qquad
\pal\pab=\delta_{\alpha\beta}\pal\;.
\end{equation}
It follows that
\begin{equation}
\pal P=P\pal=\pal\;.
\end{equation}
We can then discuss the general situation where the model-space eigenvalues
of $H_0$ are not completely degenerate.
As in the degenerate case, we have the freedom to modify the unperturbed
$P$-space Hamiltonian arbitrarily and make a compensating change in the
$P$-space part of the perturbation. Such shifts will affect the convergence
of the iteration and will determine which of the exact eigenvalues are finally
obtained. Thus we choose
\begin{equation}
PH_0'P=\sum\limits_{\alpha}\epsilon_{\alpha}\pal
\ \ {\rm and}\ \ PV'P=PVP+PH_0P-\sum\limits_{\alpha}\epsilon_{\alpha}\pal\;.
\end{equation}
This does not exclude the possibility that some degeneracy is still present,
indeed the formalism may be applied to the completely degenerate case in
which case one regains the standard LS method of the preceding section.

A general perturbative expansion has been given in ref. \cite{sonond} for the
non-degenerate case within the framework of the $\qh$-box formalism. Here we
wish to generalize the iterative scheme of sec. 2. To that end we substitute
$PH_0'P$ in eq. (15) into eq. (2) and multiply by $\pal$
from the right, yielding
\begin{equation}
QV\pal+QHQ\omega\pal-\epsilon_{\alpha}\omega\pal-\omega(PV'P+PVQ\omega)\pal=0\;,
\end{equation}
from which it follows that
\begin{equation}
\omega\pal=\frac{1}{\epsilon_{\alpha}-QHQ}QV\pal-\frac{1}{\epsilon_{\alpha}-QHQ}
\,\omega R\pal\;.
\end{equation}
Using eq. (13) and noting that $\omega P\equiv\omega$ we have a
generalization of eq. (6) to the non-degenerate case, namely
\begin{equation}
\omega=\sum\limits_{\alpha}\frac{1}{\epsilon_{\alpha}-QHQ}QV\pal
-\sum\limits_{\alpha}\frac{1}{\epsilon_{\alpha}-QHQ}\,\omega R\pal\;.
\end{equation}
Multiplying eq. (1) on the right by $\pal$, using eq. (17) and noting that
$R\equiv RP=\sum_{\alpha}R\pal$, we have a formal solution for $R$ given by
\begin{equation}
R=\sum\limits_{\alpha}\left[1+PVQ\frac{1}{\epsilon_{\alpha}-QHQ}
\,\omega\right]^{-1}\qh(\epsilon_{\alpha})\pal\;,
\end{equation}
where $\qh(\epsilon_{\alpha})$ is similar to eq. (8), namely
\begin{equation}
\hat{Q}(\epsilon_{\alpha})=PV'P+PVQ\frac{1}
{\epsilon_{\alpha}-QHQ}QVP\;.
\end{equation}

We now set up iterative equations for $\omega$ and $R$. In the most general
case the projection operators may vary according to the iteration, i.e.
$\pal\rightarrow\pal^n$.
This means that $\qh\rightarrow\qh^n$ since it is dependant on the iteration
through the
quantity $PV'P=PVP+PH_0P-\sum_{\alpha}\epa\pal^n$. We can write the iterative
equations in the form
\begin{eqnarray}
R_n&\pj=&\pj\sum\limits_{\alpha}\left[P+PVQ\frac{1}{\epsilon_{\alpha}-QHQ}\,
\omega_{n-1}\right]^{-1}\hat{Q}^{n-1}(\epsilon_{\alpha})\pal^{n-1}\;,\\
{\rm and}&&\nonumber\\
\omega_n&\pj=&\pj\sum\limits_{\alpha}\left[\frac{1}{\epsilon_{\alpha}-QHQ}QV
\pal^{n-1}-\frac{1}{\epsilon_{\alpha}-QHQ}\,\omega_{n-1}R_n\pal^{n-1}\right]\;.
\end{eqnarray}
This is just the generalization of the LS eqs. (9) and (10) to the
non-degenerate case. We define $\pal^0$ in terms of the basis states of the
original unperturbed Hamiltonian, thus
\begin{equation}
\pal^0=|\alpha\rangle\langle\alpha|\qquad{\rm where}\qquad
H_0|\alpha\rangle=\epa|\alpha\rangle\;.
\end{equation}
Taking $\omega_0=0$, we have the sequence
\begin{eqnarray}
R_1&\pj=&\pj\sum\limits_{\alpha}\qh^0(\epsilon_{\alpha})\pal^0\;,\\
R_2&\pj=&\pj\sum\limits_{\alpha}\left[P-\sum\limits_{\beta}
\qh_1(\epsilon_{\alpha},\epsilon_{\beta})\pab^0\right]^{-1}
\qh^1(\epsilon_{\alpha})\pal^1\;,\\
R_3&\pj=&\pj\sum\limits_{\alpha}\left[P-\sum\limits_{\beta}\hspace{-1.7mm}
\qh_1(\epsilon_{\alpha},\epsilon_{\beta})\pab^1
-\sum\limits_{\beta\gamma}\hspace{-1.7mm}\qh_2(\epa,\epb,\epc)\pab^0R_2
P_{\gamma}^1\right]^{-1}\hspace{-.5cm}\qh^2(\epsilon_{\alpha})\pal^2,
\end{eqnarray}
and in general $R_n$, for $n>2$, is given by
\begin{eqnarray}
R_n&\pj=&\pj\sum\limits_{\alpha}\Biggl[P-\sum\limits_{\beta}
\qh_1(\epsilon_{\alpha},\epsilon_{\beta})\pab^{n-2}
-\sum\limits_{\beta\gamma}\qh_2(\epa,\epb,\epc)\pab^{n-3}R_{n-1}
P_{\gamma}^{n-2}\cdots\nonumber\\
&\pj-&\pj\hspace{-.3cm}\sum\limits_{\beta\gamma\cdots\lambda\mu}\hspace{-.3cm}
\qh_{n-1}(\epa,\epb,\epc,\cdots\epsilon_{\lambda},\epsilon_{\mu})
\pab^0R_2P_{\gamma}^1R_3\cdots R_{n-2}P_{\lambda}^{n-3}
R_{n-1}P_{\mu}^{n-2}\Biggr]^{-1}\nonumber\\
&&\hspace{4.5cm}\qh^{n-1}(\epsilon_{\alpha})\pal^{n-1}\nonumber\\
&\pj=&\pj\sum\limits_{\alpha}\Biggl[P-\sum\limits_{\beta}
\qh_1(\epsilon_{\alpha},\epsilon_{\beta})\pab^{n-2}-\sum\limits_{m=2}^{n-1}
\sum\limits_{\beta_1\beta_2\cdots\beta_m}\hspace{-3mm}\qh_m(\epa,
\epsilon_{\beta_1},\cdots,
\epsilon_{\beta_m})\nonumber\\
&&\hspace{1cm}\times\  P_{\beta_1}^{n-m-1}\prod\limits_{k=n-m+1}^{n-1}
R_kP_{\beta_{k-n+m+1}}^{k-1}\Biggr]^{-1}\qh^{n-1}(\epa)\pal^{n-1}.
\end{eqnarray}
Here we have defined
\begin{displaymath}
\qh_1(\epsilon_1,\epsilon_2)=-PVQ\frac{1}{\epsilon_1-QHQ}
\:\frac{1}{\epsilon_2-QHQ}QVP
\end{displaymath}
$$\hspace{-3.2cm}=\cases{
               \frac{\qh(\epsilon_1)-\qh(\epsilon_2)}{\epsilon_1-\epsilon_2}
               &if $\epsilon_1\ne\epsilon_2$\cr
               \frac{d\qh(\epsilon)}{d\epsilon}\Big|_{\epsilon=\epsilon_1}
               &if $\epsilon_1=\epsilon_2$\cr}\;,\eqno(28)$$
and, in general,
\addtocounter{equation}{1}
\begin{eqnarray}
\qh_m(\epsilon_1,\epsilon_2,\cdots,\epsilon_{m+1})&\pj=&\pj(-1)^mPVQ
\frac{1}{\epsilon_1-QHQ}\:\frac{1}{\epsilon_2-QHQ}\cdots\nonumber\\
&&\qquad\qquad\qquad\quad\cdots\frac{1}{\epsilon_{m+1}-QHQ}QVP\;.
\end{eqnarray}

\subsection{Choice of Projection Operator}

The simplest choice is to keep $\pal$ fixed at its initial value, namely
\begin{equation}
\pal^n=\pal^0=|\alpha\rangle\langle\alpha|\;.
\end{equation}
This means that $PV'P=PVP$ so that $\qh^n$ is independant of $n$ and is the
same as $\qh$ of eq. (8). We refer to this as the generalized Lee-Suzuki
approach (GLS).
We can easily regain the results of sec. 2 in the completely degenerate
case where $\epa=\epsilon_0$ for all $\alpha$. Thus
$\qh_m(\epsilon_1,\epsilon_2,\cdots,\epsilon_{m+1})$ becomes simply the
$\qh_m$ of eq. (12). This means that the summations in eq. (27) are just of
the form $\sum_{\beta_i}P_{\beta_i}=P$ and since $\qh_mP=\qh_m$ we return
to eq. (11).

The other choice we consider for the projection operators is to write them
in terms of the model space wave functions obtained at each iteration.
Specifically our $n^{\rm th}$ approximation to the exact eq. (3) is
\begin{equation}
\left(\sum_{\alpha}\epa\pal^{n-1}+R_n\right)|\phi_{\alpha}^n\rangle=
E_{\alpha}^n|\phi_{\alpha}^n\rangle\;,
\end{equation}
where $E_{\alpha}^n$ is the $n^{\rm th}$ approximation to the true
eigenvalue $E_{\alpha}$. The vectors $|\phi_{\alpha}^n\rangle$, being
simply the projections of the true wave functions onto the model space, are not
orthogonal. However it is well known that their biorthogonal complements
$|\tilde{\phi}_{\alpha}^n\rangle$ can be defined such that
$\langle\tilde{\phi}_{\alpha}^n|\phi_{\beta}^n\rangle=\delta_{\alpha\beta}$.
We then define
\begin{equation}
\pal^n=|\phi_{\alpha}^n\rangle\langle\tilde{\phi}_{\alpha}^n|\;,
\end{equation}
and it is easily verified that eqs. (13) and (14) are satisfied.
It is necessary to specify which $\epa$ is associated with a given $\pal$
in eq. (31). This we do in the obvious way by ordering the unperturbed
energies $\epsilon_1<\epsilon_2<\cdots<\epsilon_d$ and the approximate
energies $E_1^n<E_2^n<\cdots<E_d^n$ and making the association with the
corresponding eigenvectors. We refer to this as the generalized Lee-Suzuki
method with self-consistent basis (SCGLS). In the completely degenerate case
$\sum_{\alpha}\epa\pal^n$ becomes $\epsilon_0\sum_{\alpha}\pal^n=\epsilon_0P$
and the approach reverts to the standard LS method by the arguments
given before.

\subsection{Evaluation of $\qh_m$}

The operator $\qh_m(\epsilon_1,\epsilon_2,\cdots,\epsilon_{m+1})$
is the basic element needed to construct the effective interaction for
the non-degenerate case and it can be expressed as a linear combination
of the standard $\qh$-boxes according to
\begin{equation}
\qh_m(\epsilon_1,\epsilon_2,\cdots,\epsilon_{m+1})=
\sum\limits_{k=1}^{m+1}C_k(\epsilon_1,\epsilon_2,\cdots,\epsilon_{m+1})
\qh(\epsilon_k)\;,
\end{equation}
where
\begin{eqnarray}
C_k(\epsilon_1,\epsilon_2,\cdots,\epsilon_{m+1})&\pj=&\pj\Bigl[
(\epsilon_k-\epsilon_1)(\epsilon_k-\epsilon_2)\cdots(\epsilon_k-\epsilon_{k-1})
(\epsilon_k-\epsilon_{k+1})\cdots\nonumber\\
&&\qquad\qquad\qquad\qquad\cdots(\epsilon_k-\epsilon_{m+1})\Bigr]^{-1}\;.
\end{eqnarray}
For the derivation of eq. (33) we have used the equality
\begin{equation}
\Bigl[(\epsilon_1-QHQ)\cdots(\epsilon_{m+1}-QHQ)\Bigr]^{-1}=
\sum\limits_{k=1}^{m+1}(-1)^m
\frac{C_k(\epsilon_1,\epsilon_2,\cdots,\epsilon_{m+1})}{\epsilon_k-QHQ},
\end{equation}
which is easily proved by induction. It should be noted that the term $PV'P$ in
the $\qh$-box (eq. (20)) gives no contribution to eq. (33) because of
the relation
\begin{equation}
\sum\limits_{k=1}^{m+1}C_k(\epsilon_1,\epsilon_2,\cdots,\epsilon_{m+1})=0\;,
\end{equation}
which can be proved directly from eq. (34).

So far we have implicitly assumed that the energies
$(\epsilon_1,\epsilon_2,\cdots,\epsilon_{m+1})$ are all different. However
in carrying out the summations we shall need to consider the case where two or
more of the energies refer to the same state and there may also be some
degeneracy present. Suppose, for instance, that $\epsilon_1=\epsilon_2$
so that the formal limit $\epsilon_1\rightarrow\epsilon_2$ leads to the
derivative as in eq. (28). This is actually evaluated by calculating
$\qh_1(\epsilon_1,\epsilon_1+\delta)$ or, in general,
$\qh_m(\epsilon_1,\epsilon_1+\delta,\epsilon_3,\cdots,\epsilon_{m+1})$,
where $\delta$ is small in comparison to $\epsilon_1$. Similarly if
$\epsilon_1=\epsilon_2=\epsilon_3$, we would use
$\qh_m(\epsilon_1-\delta,\epsilon_1,\epsilon_1+\delta,\epsilon_4,\cdots,
\epsilon_{m+1})$, and so on. Thus the fundamental quantity we need is
$\qh(\epsilon_k+p\delta)$, where $p=0,\pm1,\pm2,\ldots$ and this only
differs from the degenerate case in that here more than one value of
$\epsilon_k$ needs to be considered. Thus in applying eqs. (33) and (34)
any degeneracy should be broken by a small amount so that the appropriate
derivative is implicity evaluated.

\section{Convergence of the Iterative Solution}

We now discuss the convergence condition for the generalized LS iteration
scheme. Let $\dw_{n-1}$ and $\dw_n$ be the deviations from the exact solution
$\omega$ of eq. (18) in the $(n-1)^{\rm st}$ and $n^{\rm th}$ iterations
respectively. Thus, by definition,
\begin{eqnarray}
\omega_{n-1}&\pj=&\pj\omega+\dw_{n-1}\;,\\
\omega_n&\pj=&\pj\omega+\dw_n\;.
\end{eqnarray}
We first discuss the case where the projection operator is fixed,
{\it i.e.,} $\pal^{n}=\pal^0$.
Now setting $R_n=R+\delta R_n$, we find from eq. (1) that
$\delta R_n=PVQ\dw_n$.  Then to first order in  $\dw$, using eq. (22),
we have
\begin{equation}
\left[\epsilon_{\alpha}-(QHQ-\omega PVQ)\right]\dw_n\pal^0=
-\dw_{n-1}R\pal^0\;.
\end{equation}
Now the term $(QHQ-\omega PVQ)$ is
the $Q$-space effective Hamiltonian \cite{lees}. The eigenvalue
equation for this case can be written in the form
\begin{equation}
\langle\psi_q|(QHQ-\omega PVQ)=E_q\langle\psi_q|\;.
\end{equation}
The eigenvalues $E_q$ above and $E_p$ in eq. (3) agree with two of the
exact eigenvalues of the Hamiltonian $H$. Now dividing eq. (39) by
the operator in square brackets on the left and multiplying by
$\langle\psi_q|$ on the left and $|\phi_p\rangle$ on the right,
we obtain
\begin{equation}
\langle\psi_q|\dw_n\pal^0|\phi_p\rangle=\langle\psi_q|\dw_{n-1}
R\pal^0|\phi_p\rangle(E_q-\epa)^{-1}\hspace{-.2cm}.
\label{38}
\end{equation}
This can be simplified by using $|\tilde{\phi}\rangle$, the biorthogonal
complement to $|\phi\rangle$, and inserting
$\sum_{p'}|\phi_{p'}\rangle\langle\tilde{\phi}_{p'}|={\bf1}$ since
\begin{equation}
\langle\tilde{\phi}_{p'}|R\pal^0=(E_{p'}-\epa)
\langle\tilde{\phi}_{p'}|\pal^0\;.
\end{equation}
Performing these manipulations and summing over $\alpha$ in eq. (\ref{38}),
we finally obtain
\begin{equation}
\langle\psi_q|\dw_n|\phi_p\rangle=\sum\limits_{p'}
\langle\psi_q|\dw_{n-1}|\phi_{p'}\rangle Z^q_{p'p}\;,\label{40}
\end{equation}
where we have defined
\begin{equation}
Z^q_{p'p}=\sum\limits_{\alpha}\left(\frac{\epa-E_{p'}}{\epa-E_q}\right)
\left<\tilde{\phi}_{p'}\left|\pal^0\right|\phi_p\right>\;.\label{41}
\end{equation}
It is useful to write eq. (\ref{40}) in an obvious matrix notation as
\begin{equation}
\delta\mbox{\boldmath$\omega$}_n=\delta\mbox{\boldmath$\omega$}_{n-1}
\cdot{\bf Z}\;.
\end{equation}
Then in order that the iteration be convergent it is necessary that the norm
$\parallel\delta\mbox{\boldmath$\omega$}_n\parallel$ be smaller than
$\parallel\delta\mbox{\boldmath$\omega$}_{n-1}\parallel$ for $n$ greater than
some integer $N$. This means that a sufficient condition for convergence is
\begin{equation}
\parallel {\bf Z}\parallel<1\;.\label{43}
\end{equation}
Here we employ the Hilbert, or spectrum, norm which is defined for an arbitary
matrix ${\bf X}$ as $\parallel {\bf X}\parallel={\rm max}\sqrt{\lambda_i}$,
where
$\lambda_i$ are the eigenvalues of the matrix ${\bf X}^{\dagger}{\bf X}$. In
the completely degenerate case we see that eqs. (\ref{41}) and (\ref{43})
reduce immediately to the LS convergence condition
\begin{equation}
\left|\frac{\epsilon_0-E_p}{\epsilon_0-E_{q}}\right|<1\;.
\end{equation}
Thus eq. (\ref{43}) represents the generalization of the LS condition to the
case of a non-degenerate $P$-space.

The convergence that is obtained will depend on the choice of projection
operator. In the GLS case the condition (46) may be quite complicated, since
eq. (44) involves $\left<\tilde{\phi}_{p'}\left|\pal^0\right|\phi_p\right>=
\left<\tilde{\phi}_{p'}\left|\alpha\rangle\langle\alpha\right|\phi_p\right>$.
However,
if the states that we wish to obtain, $|\phi_p\rangle$, lie largely within the
model space the non-Hermiticity of the effective interaction will be
small (see the discussion in ref. \cite{uspl}) and
$|\tilde{\phi}_p\rangle$ will not differ greatly from $|\phi_p\rangle$.
Further if a given state $p$ contains a large component of $|\alpha\rangle$
we can expect ${\bf Z}$ to be close to a diagonal matrix. Then the convergence
condition (\ref{43}) should be satisfied if the unperturbed energy
$\epa$ is close to $E_p$, but distant from the $Q$-space eigenvalues $E_q$.

We next discuss the convergence condition for the SCGLS approach.  In this
case we have to consider the deviation $\delta \pal^{n}$ of the projection
operator defined as
\begin{equation}
\delta\pal^{n}=\pal^{n}-\pal^{\infty}\;,
\end{equation}
with
\begin{equation}
\pal^{\infty}=|\phi_\alpha\rangle\langle\tilde{\phi_\alpha}|\;,
\end{equation}
where $|\phi_{\alpha}\rangle$ is the exact eigenstate in eq. (3).
Since $R_n$ in eq. (21) can be written as
\begin{eqnarray}
R_n&\pj=&\pj PV^{\prime(n-1)}P+PVQ\omega_n\nonumber\\
&\pj=&\pj PVP+PH_0 P-\sum\limits_{\alpha}
\epsilon_{\alpha} \pal^{n-1}+PVQ\omega_n\;,
\end{eqnarray}
the deviation $\delta R_n$ is given by
\begin{equation}
\delta R_n =PVQ\delta \omega_n -\sum\limits_{\alpha}\epsilon_{\alpha}
\delta \pal^{n-1}\;.
\end{equation}
Substituting eqs. (37), (38) and (51) into eq. (22), we see that, to first
order, the terms with the deviation $\delta \pal^{n-1}$ are all canceled
and we have
\begin{equation}
[\epsilon_{\alpha}-(QHQ-\omega PVQ)]\delta \omega_n \pal^{\infty}=
-\delta \omega_{n-1} R\pal^{\infty}\;,
\end{equation}
which is just the equation obtained by replacing $\pal^0$ by $\pal^{\infty}$
in eq. (39).  Directly from eq. (52) or by replacing $\pal^0$ by
$\pal^{\infty}$ in eq. (44), we obtain the matrix $Z_{pp'}^q$ as
\begin{equation}
Z_{pp'}^q=\left(\frac{{\epsilon}_p - E_p}{{\epsilon}_p - E_q}\right)
{\delta}_{pp'}\;.
\end{equation}
We finally may say that in order that $\parallel{\bf Z}\parallel<1$ the
following condition must be satisfied:
\begin{equation}
\parallel{\bf Z}\parallel=
    \max \left|\frac{\epsilon_p - E_p}{\epsilon_p - E_q}\right|<1,\quad
p=1,2,\ldots,d,\quad q=1,2,\ldots\ \ .
\end{equation}
The above condition is satisfied if and only if the $d$ eigenvalues $E_p$
correspond to those true eigenvalues which lie nearest to the unperturbed
energies $\epsilon_p$ with $p=1,2,...,d$.

The magnitude of $||{\bf Z}||$ is equal to unity at points where
$\epsilon_p=\thalf(E_p+E_q)$, {\it i.e.}, one of the unperturbed energies is
exactly halfway between a $P$-space and a $Q$-space eigenvalue. Apart from
these points eq. (54) is obeyed, so that this iteration scheme will
converge to those eigenstates whose eigenvalues lie nearest to the
unperturbed energies $\epsilon_p$.

\section{Test Calculations}

In order to obtain some assessment of the GLS and SCGLS methods, as well as
to compare with the KK and LS methods, we need to study a model for which
exact results are readily obtained.
The model we shall use is a slightly modified version of one which was
introduced many years ago by Hoffmann {\it et al.} \cite{model}
in order to study the intruder state problem. The Hamiltonian is taken to be
$H=H_0+V$, where the unperturbed Hamiltonian matrix is
$H_0={\rm diag}(1,1,3,9)$ and the perturbation $V$ is given by
$$V=\left(\matrix{0&5x&-5x&5x\cr5x&25x&5x&-8x\cr-5x&5x&-5x&x\cr
5x&-8x&x&-5x\cr}\right)\;,$$
where $x$ is a parameter that we shall vary.
Obviously the eigenvalue problem for this model Hamiltonian
can easily be solved.

We shall take the lowest two states of $H_0$ to be our model space.
Our main concern is then to see whether an effective interaction acting
in this model space is able to reproduce any pair of exact eigenvalues. In
tables 1 and 2 we show results obtained for $x=0.05$ and 0.20. Physically these
cases differ in that for $x=0.05$ the model space states dominate the
eigenvectors of the first and second
states, whereas for $x=0.20$ they dominate the first and third eigenvectors,
{\it i.e.}, increasing the value of $x$ causes a ``level crossing" to take
place. Using the SCGLS method we show results obtained with
various choices for the $\epa$ of eq. (15). We see that as the unperturbed
energies change the solutions converge to different pairs of eigenvalues. As
predicted in sec. 4, the SCGLS yields the eigenvalues which are nearest to
the unperturbed energies. The convergence rates are reasonable in all cases,
 with accuracy to four decimal places obtained in at most 12 iterations and
usually many fewer are required. By adjusting the $\epa$ we are able to
reproduce any pair of exact eigenvalues, regardless of the order of the
eigenvalues or the magnitude of the $P$-space overlap of the
eigenvectors.

We have also carried out calculations with the GLS method where the projection
operators are defined in terms of the unperturbed eigenfunctions and do not
vary with iteration. For $x=0.20$ we were able to reproduce all pairs of
eigenvalues as with the SCGLS, however for $x=0.05$ we were not able to obtain
convergence  for three of the combinations. These were  $(E_1,E_4)$,
 $(E_2,E_4)$
and $(E_3,E_4)$. All of them involve state 4 which has a very small overlap
with the model space and the quantity $Z$ in eq. (44) depends on the overlaps
$\langle\tilde{\phi}_{p'}|\alpha\rangle$ and $\langle\alpha|\phi_p\rangle$ as
well as the energy differences  $(\epa-E_{p'})$ and $(\epa-E_q)$. Therefore
it is possible to have $\parallel {\bf Z}\parallel>1$, implying no
convergence, even though $|(\epa-E_{p'})/(\epa-E_q)|<1$.

We now turn to a comparison of the different iteration methods-- KK, LS, GLS,
SCGLS. Briefly in the KK method
\cite{emt} the effective interaction in the $n^{\rm th}$ step is given by
\begin{equation}
R_n=\sum\limits_k\qh(E^{n-1}_k)|\phi^{n-1}_k\rangle
\langle\tilde{\phi}^{n-1}_k|\;,
\end{equation}
where
\begin{equation}
(PH_0P+R_{n-1})|\phi^{n-1}_k\rangle=E^{n-1}_k|\phi^{n-1}_k\rangle\;.
\end{equation}
This iterative process sums the folded diagrams to all orders and, if it is
convergent, the states with maximium $P$-space overlap are obtained.
For $x=0.05$ the two lowest states are the ones with maximum
$P$-space overlap and for this
case we can compare the four methods. In order to make the comparison we
introduce a measure of the deviation of the calculated eigenvalues from the
exact results,
\begin{equation}
\Delta_n=\left[\sum\limits_k\left(E^n_k-E_k^{\rm exact}\right)^2
\right]^{\thalf}\;.
\end{equation}
We plot $\Delta_n$ versus the number of iterations $n$ in fig.1. Here we
have taken $\epsilon_1=1.0$ and $\epsilon_2=2.0$ for the GLS and SCGLS and a
degenerate energy of $\epsilon_0=1.5$ for LS and $\epsilon_0=1.0$ for KK.
 Of course all methods
converge, but
the SCGLS and KK techniques show the fastest convergence. Thus it appears
that the use of a self-consistent projection operator yields the most
rapid convergence. As another example, we take $x=0.20$ and reproduce the
first and third eigenvalues, which are the ones with maximum $P$-space overlap,
so that the KK, GLS and SCGLS methods can be compared. Taking
$\epsilon_0=1.0$ for KK and $\epsilon_1=0.0, \epsilon_2=5.5$ for GLS and SCGLS,
we obtain the results in
fig. 2. The convergence of the SCGLS method is markedly better than for
the other cases. In particular the GLS is rather slowly convergent and shows
an oscillatory behavior, although it does ultimately yield an accurate answer.
In fig. 3 we make a comparison between LS, GLS and SCGLS at this
value of $x$ (using  $\epsilon_0=4.0$ for LS and
$\epsilon_1=2.5,\epsilon_2=5.5$ for the other cases).
These methods converge to the second and third eigenvalues because these
lie closest to the unperturbed energies. Here the GLS and SCGLS show a better
rate of convergence than the standard LS method, with the SCGLS giving the
most rapid rate.

Finally it is of interest to examine the starting energy dependance of the
SCGLS method. We show in fig. 4 the effect of varying $\epsilon_2$ while
keeping $\epsilon_1=0.0$ for $x=0.20$. The calculations then yield the
ground-state energy and either the second or third or fourth eigenvalue
depending on $\epsilon_2$. The solid curve in fig. 4 gives the minimum number
of iterations, $n_{\rm min}$, required for $\Delta_n<10^{-4}$.
The convergence is good except at the points where
$\epsilon_2=\thalf(E_2+E_3)=4.11$ and $\epsilon_2=\thalf(E_3+E_4)=7.29$, since
here $\parallel {\bf Z}\parallel=1$. This is in agreement with the
 discussion of
sec. 4, which suggests that the error in the eigenvalue is proportional to
$||{\bf Z}||^n$, in which case
$n_{\rm min}=c/\log_{10}||{\bf Z}||$. We have plotted this quantity as the
dashed curve in fig. 4, choosing the constant $c$ to be $-5$ and calculating
$\parallel{\bf Z}\parallel$ with the
$Q$-space eigenstate nearest to $\epsilon_2$. As can be seen the dashed curve
agrees very well with the exact result.

\section{Concluding Remarks}

We have derived an iteration method (GLS) for effective interactions as a
generalization of the Lee-Suzuki method so that one can apply it to a system
with arbitrary non-degenerate unperturbed energies.  It has been proved that
the
iterative solution can be constructed in the framework of the $\hat{Q}$-box
formalism of Kuo {\it et al.} [3,5] in spite of the non-degeneracy of the
unperturbed energies.  The convergence in the GLS method depends on the
overlaps of the exact $P$-space eigenstates and the unperturbed states.
If the overlaps are small, the GLS approach does not always converge.

The GLS method has further been generalized so that different
projection operators can be used in each step of iteration.  It has been
shown that if we construct the projection operators using the $P$-space
eigenstates determined self-consistently at each iteration step, the
convergence is governed
only by energy ratios of the differences between true and unperturbed
energies and there is, in general, a unique way of distributing the true
eigenvalues between the sets $E_p$ and $E_q$, where $E_p$ denotes the
eigenvalues obtained from
the $P$-space eigenvalue equation.  This implies that, by
appropriate choice of the unperturbed energies, the $d$ eigenvalues obtained
from the effective interaction can be tuned to generate any
subset of the true eigenvalues.  We have verified this theoretical prediction
in the model  calculations.  We have observed some divergent cases for the
GLS, but the use of  self-consistent projection operators
in the SCGLS renders the iterations convergent.  It also accelerates the
convergence compared to the usual L-S method.

\vskip.5cm
This work was initiated during the authors' visit to the Nuclear
Theory Institute at the University of Washington and we thank the
Institute for the hospitality extended to us.
Partial support from the US Department of Energy under contracts
DE-FG02-88ER40388 and DE-FG02-87ER40328 is gratefully acknowledged.

\newpage
\centerline{Table 1. Convergence of the eigenvalues obtained with the
SCGLS}
\centerline{method for the model Hamiltonian with $x=0.05$}
\begin{center}
\begin{tabular}{|cc|ccccc|c|}\hline
&&\multicolumn{5}{|c|}{Number of iterations}&Exact\\
$\epsilon_1$&$\epsilon_2$&2&4&6&8&10&eigenvalues\\  \hline
1.0&2.0&0.8915&c&c&c&c&$E_1$\\
   &   &2.2177&2.2158&c&c&c&$E_2$\\[1mm]
1.0&3.0&0.8966&0.8904&c&c&c&$E_1$\\
   &   &2.6998&2.8553&2.8619&c&c&$E_3$\\[1mm]
1.0&9.0&0.9570&0.8900&0.8904&c&c&$E_1$\\
   &   &6.8904&8.7779&c&c&c&$E_4$\\[1mm]
2.0&3.0&0.8681&0.5324&2.5126&2.2268&2.2161$^*$&$E_2$\\
   &   &2.7209&2.8541&2.8466&2.8621&c&$E_3$\\[1mm]
2.0&9.0&0.9677&1.2027&2.0943&2.2106&2.2155$^{**}$&$E_2$\\
   &   &6.8689&8.7827&8.7816&c&c&$E_4$\\[1mm]
3.0&9.0&1.6722&3.6832&2.8731&2.8625&c&$E_3$\\
   &   &7.6619&8.7861&c&c&c&$E_4$\\\hline
\end{tabular}
\end{center}
\noindent The notation c indicates convergence to four decimal places. The
exact
eigenvalues here are $E_1=0.8905,\ E_2=2.2157,\ E_3=2.8622$ and $E_4=8.7817$
and the overlaps of the exact eigenfunctions with the model space are 0.973,
0.902, 0.121 and 0.004 respectively.

\noindent$^*$Convergence to four  decimal places is obtained after 12
iterations.

\noindent$^{**}$Convergence to four decimal places is obtained after 11
iterations.
\newpage
\centerline{Table 2. Convergence of the eigenvalues obtained with the
SCGLS}
\centerline{method for the model Hamiltonian with $x=0.20$}
\begin{center}
\begin{tabular}{|cc|cccc|c|}\hline
&&\multicolumn{4}{|c|}{Number of iterations}&Exact\\
$\epsilon_1$&$\epsilon_2$&2&4&6&8&eigenvalues\\  \hline
0.0&2.5&$-$3.0221&$-$0.2385&c&c&$E_1$\\
   &   &\x4.9933&\x2.5150&\x2.5793&c&$E_2$\\[1mm]
0.0&5.5&$-$0.1608&c&c&c&$E_1$\\
   &   &\x5.6553&c&c&c&$E_3$\\[1mm]
0.0&9.0&$-$0.1095&$-$0.1495&c&c&$E_1$\\
   &   &\x8.8437&c&c&c&$E_4$\\[1mm]
2.5&5.5&\x2.0849&\x2.5784&c&c&$E_2$\\
   &   &\x5.4309&c&c&c&$E_3$\\[1mm]
2.5&9.0&\x2.5823&c&c&c&$E_2$\\
   &   &\x8.9173&c&c&c&$E_4$\\[1mm]
5.5&9.0&\x1.4692&\x6.3648&\x5.6463&c&$E_3$\\
   &   &\x8.9215&\x8.9254&c&c&$E_4$\\\hline
\end{tabular}
\end{center}
\noindent The notation c indicates convergence to four decimal places. The
exact
eigenvalues here are $E_1=-0.1496,\ E_2=2.5794,\ E_3=5.6451$ and $E_4=8.9251$
and the overlap of the exact eigenfunctions with the model space are 0.715,
0.295, 0.759 and 0.231 respectively.
\newpage
\begin{center}
{\large{\bf Figure Captions}}
\end{center}

\noindent Figure 1.\ \ \ The error in the calculated eigenvalues as a
function of the number of iterations. Smooth curves are drawn through the
points to guide the eye. The curves are labelled by the method employed to
solve the model Hamiltonian problem with $x=0.05$.

\noindent Figure 2.\ \ \ As for fig. 1, but with $x=0.20$. Here the
calculations are converging to the first and third eigenvalues.

\noindent Figure 3.\ \ \ As for fig. 1, but with $x=0.20$. Here the
calculations are converging to the second and third eigenvalues.

\noindent Figure 4.\ \ \ Number of iterations required for an accuracy of
one part in $10^4$ as a function of the unperturbed energy, $\epsilon_2$,
with $x=0.20$. The full curve is obtained with the SCGLS method and the
dashed curve gives the results expected from the convergence analysis, namely
$-5/\log_{10}\parallel{\bf Z}\parallel$.
\end{document}